\documentclass[pra,twocolumn,showpacs,preprintnumbers,amsmath,amssymb]{revtex4}

\usepackage{graphicx}
\usepackage{dcolumn}
\usepackage{bm}
\newcommand{\bra}[1]{\langle#1|}

\newcommand{\scal}[2]{\langle#1|#2\rangle}\newcommand{\ket}[1]{|#1\rangle}

\begin{document}

\preprint{APS/123-QED}

\title{Efficient generation of $N$-photon generalized binomial states in a cavity}

\author{Rosario Lo Franco}
 \email{lofranco@fisica.unipa.it}
 \homepage{http://www.fisica.unipa.it/~lofranco}
\author{Giuseppe Compagno}
  \author{Antonino Messina}
 \author{Anna Napoli}

\affiliation{%
CNISM and Dipartimento di Scienze Fisiche ed Astronomiche,
Universit\`a di Palermo, via Archirafi 36, 90123 Palermo, Italy}

\date{\today}

\begin{abstract}
Extending a previous result on the generation of two-photon generalized binomial field states, here we propose an efficient scheme to generate with high-fidelity, in a single-mode high-$Q$ cavity, $N$-photon generalized binomial states with a maximum number of photons $N>2$. Besides their interest for classical-quantum border investigations, we discuss the applicative usage of these states in realizing universal quantum computation, describing in particular a scheme that performs a controlled-NOT gate by dispersive interaction with a control atom. We finally analyze the feasibility of the proposed schemes, showing that they appear to be within the current experimental capabilities.
\end{abstract}


\maketitle

\section{\label{intro}Introduction}
Nonclassical states of electromagnetic field are an important theoretical and experimental resource in quantum optics, being they relevant for fundamental aspects of quantum mechanics and also for applications in quantum information and computation \cite{zeilinger1998RevModPhys,nielsenchuang,haroche2006book}. ``Quantum field state engineering'' has thus become of interest; cavity quantum electrodynamics (CQED) has been successful as a suitable framework to this purpose \cite{raimond2001RevModPhys,haroche2006book}. CQED schemes have been proposed to generate quantum field states using the interaction of consecutive atoms with a high-$Q$ cavity \cite{meystre,vogel1993PRL}. A two-photon Fock state was also generated and probed \cite{bertet2002PRL}.

An important class of nonclassical electromagnetic field states is that of the $N$-photon generalized binomial states ($N$GBSs) \cite{stoler1985OptActa,vidiella1994PRA}. Recently, $N$GBSs have been object of a new interest because they can be utilized as reference states in schemes to measure the canonical phase of quantum electromagnetic field states \cite{pregnell2002PRL,pregnell2003PRA}. It has been recently pointed out \cite{lofranco2008EPJ} that the $N$GBSs are the exact electromagnetic correspondent of the well-known coherent atomic states \cite{arecchi1972PRA}. This correspondence suggests that $N$GBSs could be naturally produced by interactions between two-level atoms and quantum electromagnetic field. In fact, it has been shown that it is possible to efficiently generate both single $2$GBSs and their quantum superpositions \cite{lofranco2006PRA,lofranco2007PRA}. Entangled 2GBSs can also be obtained, in two spatially separated cavities, by standard resonant atom-cavity interactions \cite{lofranco2006OpenSys,lofranco2007OptSpect}.

In order to investigate the quantum-classical border it appears important to consider quantum superpositions of orthogonal states that approach the classical limit. Within the electromagnetic field states, Fock (number) states present nonclassical features also at high $N$ while coherent states, although suitable for the classical limit, are never exactly orthogonal. On the other hand, $N$GBSs are intermediate between number and coherent states because for large $N$ they approach coherent states and have thus a classical limit, while given an arbitrary $N$GBS it is always possible to find an orthogonal one \cite{lofranco2005PRA}. Moreover, they present nonzero field expectation values. These characteristics allow to construct quasi-classical superpositions of orthogonal $N$GBSs. These superpositions could also find applications in quantum information. For this, it is crucial to find realistic efficient schemes to generate $N$GBSs with $N>2$. This constitutes one of the main issues of this work. This is here accomplished by extending a CQED generation procedure already proposed and limited to a maximum number of photons $N=2$ \cite{lofranco2006PRA,lofranco2007PRA}. A second purpose of this paper is to discuss the applicative role played by $N$GBSs (with $N>2$) for realizing universal quantum computation within the CQED framework. In particular, we show that it is possible to realize a controlled-NOT (CNOT) gate that relies on the possibility to prepare a $N$GBS in a cavity and then to control its dispersive interaction with a Rydberg atom.

The paper is organized as follows. In Sec.~\ref{GBSandHamiltonian} we recall the definition of a $N$GBS and the Hamiltonian model which describes the dynamics of our system. In Sec.~\ref{genbin} we illustrate our efficient generation scheme of a $N$GBS in a cavity. In Sec.~\ref{quantumcomputation} we show the realization of a CNOT gate by means of $N$GBSs and a control atom. In Sec.~\ref{experimentalfeasibility} we analyze the experimental feasibility of the proposed schemes and in Sec.~\ref{conclu} we give our conclusions.

\section{\label{GBSandHamiltonian}Generalized binomial states and Hamiltonian model}
The normalized $N$-photon generalized binomial state ($N$GBS) is defined as \cite{stoler1985OptActa}
\begin{equation}\label{bin}
\ket{N,p,\phi}=\sum_{n=0}^Nb_n^{(N)}\left[p^{n}(1-p)^{N-n}\right]^{1/2}e^{in\phi}\ket{n},
\end{equation}
where $0\leq p\leq1$ is the probability of single photon occurrence, $\phi$ the mean phase \cite{vidiella1994PRA} and $b_n^{(N)}$ are the binomial coefficients
\begin{equation}\label{bincoeff}
b_n^{(N)}={N\choose n}^{1/2}=\left[\frac{N!}{(N-n)!\ n!}\right]^{1/2}.
\end{equation}
In the case $\phi=0$, the properties of this state \cite{stoler1985OptActa,dattoli1987JOptSoc,vidiella1994PRA,vidiella1995JModOpt}, also during interaction with atoms \cite{jos1,jos2}, have been studied. The $N$GBS of Eq.~(\ref{bin}) becomes the vacuum state $\ket{0}$ when $p=0$ and the number state $\ket{N}$ when $p=1$. For $N\rightarrow\infty$ and $p\rightarrow0$, so that $Np=\textrm{cost}\equiv|\alpha|^2$, the $N$GBS becomes the coherent state $\ket{|\alpha|e^{i\phi}}$. Thus, $N$GBSs are intermediate between the number and the coherent state. Finally, the states $\ket{N,p,\phi}$ and $\ket{N,1-p,\pi+\phi}$ are orthogonal \cite{lofranco2005PRA}.

The resonant interaction of a two-level atom with a single-mode high-$Q$ cavity is described by the usual Jaynes-Cummings Hamiltonian \cite{jaynescummings}
\begin{equation}
H_{JC}=\hbar\omega\sigma_{z}/2+\hbar\omega a^{\dag}a+i\hbar
g(\sigma_{+}a-\sigma_{-}a^{\dag}) \label{H}
\end{equation}
where $\omega$ is the cavity field mode, $g$ the atom-field coupling constant, $a$ and $a^{\dag}$ the field annihilation and creation operators and $\sigma_{z}=\ket{\uparrow}\bra{\uparrow}-\ket{\downarrow}\bra{\downarrow}$, $\sigma_{+}=\ket{\uparrow}\bra{\downarrow}$, $\sigma_{-}=\ket{\downarrow}\bra{\uparrow}$ the pseudo-spin atomic operators, $\ket{\uparrow}$ and $\ket{\downarrow}$ being respectively the excited and ground state of the two-level atom. The Hamiltonian $H_{JC}$ generates the transitions \cite{scully1997book}
\begin{eqnarray}
\ket{\uparrow n}&\rightarrow&\cos(g\sqrt{n+1}t)\ket{\uparrow n}-\sin(g\sqrt{n+1}t)\ket{\downarrow n+1}\nonumber\\
\ket{\downarrow n}&\rightarrow&\cos(g\sqrt{n}t)\ket{\downarrow
n}+\sin(g\sqrt{n}t)\ket{\uparrow n-1},\label{evo}
\end{eqnarray}
where $\ket{\uparrow}\ket{n}\equiv\ket{\uparrow n}$, $\ket{\uparrow}\ket{n}\equiv\ket{\downarrow n}$, $a^\dag a\ket{n}=n\ket{n}$ and $t$ is the atom-cavity interaction time.

\section{\label{genbin}Efficient generation scheme of a $N$GBS with $N>2$ in a cavity}
In this section we describe an efficient scheme to generate $N$GBSs in a single-mode high-$Q$ cavity.

A conditional scheme by which $N$GBSs are generated in a cavity has been already proposed \cite{moussa1998PLA}. This scheme utilizes the resonant interaction of $N$ consecutive monokinetic two-level atoms with the single-mode cavity, initially in the vacuum state. This scheme requires that all the $N$ atoms are detected in their ground state, and the success generation probability is of the order of $1/2^N$ \cite{vogel1993PRL,moussa1998PLA}. Thus, the efficiency rapidly decreases as $N$ increases.

Alternatively, an efficient, quasi-deterministic scheme has been presented, that exploits the resonant interaction of two consecutive two-level atoms with the single-mode cavity field, to generate a 2GBS in a cavity \cite{lofranco2006PRA}. The essence of this scheme is the utilization of different interaction times for the two atoms, that allows to obtain the desired cavity state without the necessity of a final atomic state detection. However, exploitation of this scheme is limited to a maximum number of photons $N=2$ and it is not generalizable to a larger value of $N$. Our aim is thus to find a scheme that efficiently generates the $N$GBS defined in Eq.~(\ref{bin}) for $N>2$.

\begin{figure}
\includegraphics[width=0.46\textwidth]{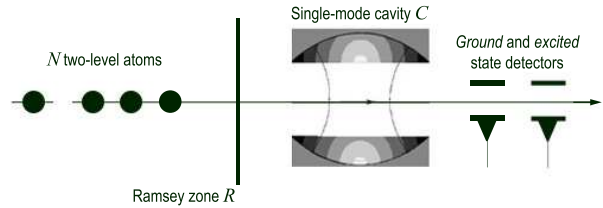}
\caption{\footnotesize Experimental scheme to generate a $N$GBS in a high-$Q$ cavity. The Ramsey zone $R$ prepares each atom in a desired superposition of its two level.}\label{figatomcav}
\end{figure}

The experimental setup of our general scheme is sketched in Fig.~\ref{figatomcav}.
$N$ consecutive two-level atoms resonantly interact, one at a time, first with a Ramsey zone \cite{ramsey1985book}, where they are prepared in a superposition of its two levels, and successively with the cavity at the beginning in the vacuum state. Finally, they are detected by field ionization detectors. The probability to generate the desired $N$GBS is thus given by the probability to detect all the atoms in the ground state. We shall not consider the free evolution times of atoms and cavity field, because they can be set at zero in an experiment by appropriately adjusting the atomic separation and the distances between the interaction zones \cite{hagley1997PRL}. As suggested by the previous scheme for the 2GBS generation \cite{lofranco2006PRA}, the idea is to exploit suitable atom-cavity interaction times, different for each atom. These different interaction times can be experimentally obtained by selecting either different velocities for each atom or, more easily, the same velocity for all the atoms (``monokinetic atomic beam'') and then applying a Stark shift inside the cavity for a time such as to have the desired resonant interaction time \cite{hagley1997PRL}. The initial atomic velocity can be selected by laser induced atomic pumping \cite{hagley1997PRL}. The typical cavity quality factor $Q$ must be sufficiently high, as will be later shown to be the case, to allow the cavity field not to significatively decay during the time necessary for the crossing of $N$ atoms, with $N$ sufficiently large.

\subsection{\label{generalprocedure}The general procedure}
In our scheme the $k$-th two-level atom ($k=1,\ldots,N$) is prepared by the Ramsey zone in a superposition of its levels given by
\begin{equation}
\ket{\chi_k}=\sqrt{p}\ket{\uparrow}+e^{i\varphi_k}\sqrt{1-p}\ket{\downarrow},\quad(k=1,\ldots,N)\label{chik}
\end{equation}
where $0\leq p\leq1$ is equal for all the atoms and $\varphi_k$ is the relative phase between the two atomic levels which will be related to the mean phase $\phi$ of the $N$GBS to be generated. The values of $p$ and $\varphi_k$ can be fixed by adjusting the Ramsey zone settings, i.e. the classical field (laser) amplitude and the atom-field interaction time.

Let us consider the $k$-th step of the generation procedure and suppose that, before the injection of the $k$-th atom, the $(k-1)$-th atom has left the cavity in a state of the form
\begin{equation}\label{psik-1state}
\ket{\psi_{k-1,p,\phi}}=\frac{1}{\mathcal{N}_{k-1}}\sum_{n=0}^{k-1}c_n^{(k-1)}\sqrt{p^{n}(1-p)^{k-1-n}}
e^{in\phi}\ket{n},
\end{equation}
where ${\cal N}_{k-1}$ is a normalization constant and the coefficients $c_n^{(k-1)}$ are real. Note that for $k=1$ this state reduces to the vacuum state $\ket{0}$. Before the $k$-th atom enters the cavity, the total atom-cavity state is thus factorized in $\ket{\chi_k}\ket{\psi_{k-1,p,\phi}}$. The $k$-th atom then resonantly interacts with the cavity for a time $T_k$. Using Eqs.~(\ref{evo}), (\ref{chik}) and (\ref{psik-1state}) we find that, after the interaction time $T_k$, the total atom-cavity state becomes, apart from an unimportant global phase factor,
\begin{eqnarray}\label{Psik}
\ket{\Psi_k(T_k)}&=&\frac{1}{\mathcal{N}_{k-1}}\sum_{n=1}^ka_n^{(k)}\sqrt{p^{n}(1-p)^{k-n}}e^{in\phi}\ket{n-1}\ket{\uparrow}
\nonumber\\
&+&\sum_{n=0}^kc_n^{(k)}\sqrt{p^{n}(1-p)^{k-n}}e^{in\phi}\ket{n}\ket{\downarrow},
\end{eqnarray}
where
{\setlength\arraycolsep{1pt}\begin{eqnarray}\label{coefficientsaandc}
a_n^{(k)}&=&c_{n-1}^{(k-1)}\cos(g\sqrt{n}T_k)+e^{i\Phi_k}c_{n}^{(k-1)}\sin(g\sqrt{n}T_k),\nonumber\\
c_n^{(k)}&=&e^{i\Phi_k}c_n^{(k-1)}\cos(g\sqrt{n}T_k)-c_{n-1}^{(k-1)}\sin(g\sqrt{n}T_k),
\end{eqnarray}}with $\Phi_k=\phi+\varphi_k$. We now suppose that the state $\ket{\psi_{k-1,p,\phi}}$ of Eq.~(\ref{psik-1state}) is exactly the $(k-1)$GBS $\ket{k-1,p,\phi}$ defined in Eq.~(\ref{bin}), for which ${\cal N}_{k-1}=1$ and $c_n^{(k-1)}=b_n^{(k-1)}$. In this case, if $a_n^{(k)}=0$ for each $n=1,\ldots,k$, that is the $k$ conditions
\begin{equation}
b_{n-1}^{(k-1)}\cos(g\sqrt{n}T_k)\pm b_{n}^{(k-1)}\sin(g\sqrt{n}T_k)=0,\label{eqcoeff}
\end{equation}
are simultaneously satisfied, it is readily seen from Eq.~(\ref{Psik}) that the atom would deterministically exit the cavity in the ground state $\ket{\downarrow}$ and, as shown in Appendix~\ref{binomialcoeffcondition}, we would have $c_n^{(k)}=b_n^{(k)}$. The sign ``$+$'' (``$-$'') in Eq.~(\ref{eqcoeff}) corresponds to set $\Phi_k=0\ (\pi)$ in Eq.~(\ref{coefficientsaandc}). Thus, the resulting cavity field state would be just the $k$GBS $\ket{k,p,\phi}$ of Eq.~(\ref{bin}), provided that negative (positive) values of the functions $\sin(g\sqrt{n}T_k)$ are chosen. Note that the appropriate relative phase $\varphi_k$ of the prepared atomic state $\ket{\chi_k}$ is fixed by the condition $\Phi_k=0\ (\pi)$ as
\begin{equation}\label{atomrelativephase}
\varphi_k+\phi=0\ (\pi),
\end{equation}
and it depends on the phase $\phi$ of the previous cavity state. While for the case $k=1$ the only equality of Eq.~(\ref{eqcoeff}) can be exactly satisfied by a single value of $T_1$, unfortunately, for each $k>1$ the $k$ conditions of Eq.~(\ref{eqcoeff}) cannot be simultaneously satisfied by a single value of $T_k$. However, it can be shown by a numerical analysis that a value of the interaction time $T_k$ exists such that the $k$ equalities of Eq.~(\ref{eqcoeff}) are with good approximation satisfied and the resulting cavity state, after measuring the $k$-th atom in the ground state, is with high fidelity the $k$GBS. The effective cavity state resulting from this $k$-th step of the procedure has thus the form
\begin{equation}\label{psikstate}
\ket{\psi_{k,p,\phi}}=\frac{1}{\mathcal{N}_k}\sum_{n=0}^{k}c_n^{(k)}\left[p^{n}(1-p)^{k-n}\right]^{1/2}e^{in\phi}\ket{n},
\end{equation}
as seen from Eq.~(\ref{Psik}), and it is obtained with probability $P_k(p)=\mathcal{N}_k^2/\mathcal{N}_{k-1}^2$ given by
{\setlength\arraycolsep{1pt}\begin{eqnarray}\label{probabilityk}
P_k(p)&=&\frac{\displaystyle1-\sum_{n=0}^k[(b_n^{(k)})^2-(c_n^{(k)})^2]p^n(1-p)^{k-n}}{\displaystyle
1-\sum_{n=0}^{k-1}[(b_n^{(k-1)})^2-(c_n^{(k-1)})^2]p^n(1-p)^{k-1-n}}.\nonumber\\
\end{eqnarray}}The procedure ends successfully when all the $N$ atoms have been detected in the ground state, so that the final cavity state is $\ket{\psi_{N,p,\phi}}$ of Eq.~(\ref{psikstate}) with $k=N$, generated with probability
\begin{equation}\label{ProbabilityN}
\mathcal{P}_N(\ket{\downarrow_N}|\ket{\downarrow_{N-1}}|\cdots|\ket{\downarrow_1})=\prod_{k=1}^NP_k(p).
\end{equation}
In order to evaluate how much the generated cavity state $\ket{\psi_{N,p,\phi}}$ is near to the desired $N$GBS $\ket{N,p,\phi}$, we calculate the fidelity $\mathcal{F}_N(p)=|\scal{N,p,\phi}{\psi_{N,p,\phi}}|^2$ whose explicit form is
\begin{equation}\label{fidelityN}
F_N(p)=\frac{\displaystyle\left[\sum_{n=0}^Nb_n^{(N)}c_n^{(N)}p^n(1-p)^{N-n}\right]^2}{\displaystyle
\sum_{n=0}^{N}(c_n^{(N-1)})^2p^n(1-p)^{N-n}}.
\end{equation}
It is evident that the more the effective coefficients $c_n^{(N)}$ are near to the binomial coefficients $b_n^{(N)}$, the more both the generation probability and the fidelity are near to one.

Among the possible ways by which the numerical analysis can be accomplished, we choose to look for an interaction time $T_k$ that is solution of the condition $n=k-1$ of Eq.~(\ref{eqcoeff}), taking into account that the typical CQED experimental conditions limit the interaction times $T$ inside the range $10^{-1}\leq gT\leq10^2$ \cite{haroche2006book}. This procedure will be shown to be efficient. For each $k>1$ the binomial coefficients $b_n^{(k-1)}$ appearing in Eq.~(\ref{eqcoeff}) will be then substituted by the effective coefficients $c_n^{(k-1)}$ resulting from the previous $(k-1)$-th state $\ket{\psi_{k-1,p,\phi}}$ of the procedure. In the following, we give the values of the suitable interaction times for each $1\leq k\leq N$, reporting the generation probabilities and the fidelity of the target $N$GBSs. The first two steps of the procedure are only intermediate steps to generate $N$GBSs with $N$ larger than two.

As said before, the cavity is initially taken in the vacuum state $\ket{0}$. The atom 1 is prepared in the superposition $\ket{\chi_1}$ of Eq.~(\ref{chik}) with an arbitrary relative phase $\varphi_1$ that will be related to the mean phase of the final generated $N$GBS. The evolved atom-cavity state at time $T_1$ is readily obtained by Eq.~(\ref{Psik}) and has the form
\begin{eqnarray}
\ket{\Psi_1(T_1)}&=&\cos(gT_1)\sqrt{p}\ket{0}\ket{\uparrow}+\big(e^{i\varphi_1}\sqrt{1-p}\ket{0}\nonumber\\
&-&\sqrt{p}\sin(gT_1)\ket{1}\big)\ket{\downarrow}.
\end{eqnarray}
We choose the interaction time
\begin{equation}
T_1=3\pi/2g,\label{T1}
\end{equation}
so that the final total state, after the atom 1 exits the cavity, is factorized as $\ket{\Psi_1(T_1)}=e^{i\varphi_1}\ket{1,p,-\varphi_1}\ket{\downarrow}$, where the atom is in the ground state and the cavity field state is in the 1GBS given by Eq.~(\ref{bin}) with $N=1$ and $\phi=-\varphi_1$.

The atom 2 is then prepared in the superposition $\ket{\chi_2}$ of Eq.~(\ref{chik}) with relative phase
\begin{equation}
\varphi_2=\varphi_1,
\end{equation}
and it interacts with the cavity for a time
\begin{equation}
T_2=7\pi/4g.\label{T2b}
\end{equation}
This gives an intermediate cavity state $\ket{\psi_{2,p,\phi}}$ of the form of Eq.~(\ref{psikstate}) with
$\phi=-\varphi_1$ and coefficients $c_n^{(2)}$ given by
\begin{equation}
c_0^{(2)}=1,\ c_1^{(2)}=\sqrt{2},\ c_2^{(2)}\approx-(1-3.114\times10^{-3}).\label{coeffcn2}
\end{equation}
At this stage, the state $\ket{\psi_{2,p,\phi}}$ only has the role of intermediate state for generating the $N$GBS with $N>2$. The probability to obtain this intermediate cavity state, equal to the probability to detect the second atom in the ground state, is
\begin{equation}
\mathcal{P}_2(p)\approx1-6.22\times10^{-3}p^{2},\quad(0\leq p\leq1).\label{Prob2}
\end{equation}

In the next step, the atom 3 is prepared in the superposition $\ket{\chi_3}$ given by Eq.~(\ref{chik}) with relative phase \begin{equation}
\varphi_3=\varphi_1,
\end{equation}
and its interaction time with the cavity is
\begin{equation}
T_3=\frac{1}{g\sqrt{2}}[\arctan(c_1^{(2)}/|c_2^{(2)}|)+5\pi],\label{T3}
\end{equation}
where the coefficients $c_1^{(2)},c_2^{(2)}$ are given in Eq.~(\ref{coeffcn2}). This interaction time is such that $gT_3\approx11.784$. This gives a resulting cavity state $\ket{\psi_{3,p,\phi}}$ of the form of Eq.~(\ref{psikstate}) with $k=N=3$, mean phase $\phi=-\varphi_1$ and the coefficients $c_n^{(3)}$ listed in Tab.~\ref{table}. The probability $P_3(p)$ to detect the the third atom in the ground state is found from Eq.~(\ref{probabilityk}) and it has the value
\begin{equation}
P_3(p)\approx1-8\times10^{-2}\frac{p(1-p)^2}{1-6\times10^{-3}p^2},\label{pro3}
\end{equation}
which is inside the range $1-10^{-2}\leq P_3(p)\leq1$ for any value of $p$. The probability to generate the cavity state $\ket{\psi_{3,p,\phi}}$ is then, from Eq.~(\ref{ProbabilityN}),
\begin{equation}
\mathcal{P}_3(\ket{\downarrow_3}|\ket{\downarrow_2}|\ket{\downarrow_1})=P_3(p)\cdot P_2(p)\cdot1\geq1-10^{-2},
\end{equation}
much higher than the correspondent generation probability of precedent conditional schemes $\sim1/8$ \cite{vogel1993PRL,moussa1998PLA}. In order to estimate how much the resulting state $\ket{\psi_{3,p,\phi}}$ is near to the ``target'' 3GBS $\ket{3,p,\phi}$, we calculate the fidelity $\mathcal{F}_3(p)$ of Eq.~(\ref{fidelityN}), which is found to tend to one for $p\rightarrow0$ and for $p=1/2$ takes the value
\begin{equation}
\mathcal{F}_3(1/2)\approx1-3.9\times10^{-5}.
\end{equation}
Thus, the state $\ket{\psi_{3,p,\phi}}$ can be effectively identified with the 3GBS $\ket{3,p,\phi}$ of Eq.~(\ref{bin}).

\begin{table}
\footnotesize
\caption{\label{table}\footnotesize Comparison between the coefficients $c_n^{(N)}$ of the generated cavity state $\ket{\psi_{N,p,\phi}}$ and the effective binomial coefficients $b_n^{(N)}=\sqrt{N!/(N-n)!n!}$, where $N$ is the maximum number of photons of the cavity field state and $n=0,\ldots,N$. We list $c_n^{(N)}=b_n^{(N)}(1-\delta_n^{(N)})$ for $3\leq N\leq10$.}
\begin{ruledtabular}
\begin{tabular}{ccc}
$n\footnotemark[1]$ &$c_n^{(3)}$&$c_{n}^{(4)}$\\
\hline 0&1&1\\
1& $\sqrt{3}(1-1.391\times10^{-2})$&$2(1-1.325\times10^{-2})$\\
2& $\sqrt{3}(1-0.981\times10^{-4}$)&$\sqrt{6}(1-1.041\times10^{-2}$)\\
3& $1-3.167\times10^{-3}$&$2(1-1.525\times10^{-3}$)\\
4&- &$1-6.720\times10^{-2}$\\
\end{tabular}
\end{ruledtabular}
\\${}$\\${}$\\
\begin{ruledtabular}
\begin{tabular}{ccc}
$n\footnotemark[1]$ &$c_{n}^{(5)}$& $c_{n}^{(6)}$\\
\hline 0&1&1\\
1& $\sqrt{5}(1-1.533\times10^{-2}$)&$\sqrt{6}(1-1.816\times10^{-2}$)\\
2& $\sqrt{10}(1-1.846\times10^{-2}$)&$\sqrt{15}(1-2.618\times10^{-2}$)\\
3& $\sqrt{10}(1-1.183\times10^{-2}$)& $\sqrt{20}(1-2.442\times10^{-1}$)\\
4& $\sqrt{5}(1-1.434\times10^{-2}$)& $\sqrt{15}(1-1.812\times10^{-2}$)\\
5& $1-1.100\times10^{-1}$& $\sqrt{6}(1-2.963\times10^{-2}$)\\
6&- & $1-1.424\times10^{-1}$\\
\end{tabular}
\end{ruledtabular}
\\${}$\\${}$\\
\begin{ruledtabular}
\begin{tabular}{ccc}
$n$ &$c_n^{(7)}$&$c_{n}^{(8)}$\\
\hline 0&1&1\\
1& $\sqrt{7}(1-2.107\times10^{-2}$)& $\sqrt{8}(1-2.384\times10^{-2}$)\\
2& $\sqrt{21}(1-3.334\times10^{-2}$)& $\sqrt{28}(1-3.984\times10^{-2}$)\\
3& $\sqrt{35}(1-3.658\times10^{-2}$)& $\sqrt{56}(1-4.761\times10^{-2}$)\\
4& $\sqrt{35}(1-3.213\times10^{-2}$)& $\sqrt{70}(1-4.731\times10^{-2}$)\\
5& $\sqrt{21}(1-2.684\times10^{-2}$)&$\sqrt{56}(1-4.101\times10^{-2}$)\\
6&$\sqrt{7}(1-4.490\times10^{-2}$)&$\sqrt{28}(1-3.664\times10^{-2}$)\\
7& $1-1.686\times10^{-1}$&$\sqrt{8}(1-5.950\times10^{-2}$)\\
8&- &$1-1.908\times10^{-1}$\\
\end{tabular}
\end{ruledtabular}
\\${}$\\${}$\\
\begin{ruledtabular}
\begin{tabular}{ccc}
$n$ &$c_{n}^{(9)}$& $c_{n}^{(10)}$\\
\hline 0&1&1\\
1& $3(1-2.640\times10^{-2}$) & $\sqrt{10}(1-2.874\times10^{-2}$)\\
2& $6(1-4.578\times10^{-2}$)& $\sqrt{45}(1-5.101\times10^{-2}$)\\
3& $\sqrt{84}(1-5.753\times10^{-2}$) & $\sqrt{120}(1-6.633\times10^{-2}$)\\
4& $\sqrt{126}(1-6.156\times10^{-2}$) & $\sqrt{210}(1-7.446\times10^{-2}$)\\
5&$\sqrt{126}(1-5.835\times10^{-2}$) &$\sqrt{252}(1-7.528\times10^{-2}$)\\
6&$\sqrt{84}(1-4.052\times10^{-2}$) &$\sqrt{210}(1-6.956\times10^{-2}$)\\
7& $6(1-4.678\times10^{-2}$) & $\sqrt{120}(1-6.034\times10^{-2}$)\\
8& $3(1-7.320\times10^{-2}$) & $\sqrt{45}(1-5.692\times10^{-2}$)\\
9& $1-2.100\times10^{-1}$& $\sqrt{10}(1-8.608\times10^{-2}$)\\
10&- & $1-2.268\times10^{-1}$\\
\end{tabular}
\end{ruledtabular}
\footnotetext[1]{Note that $\forall N$ it results
$c_0^{(N)}=b_0^{(N)}=1$.}
\end{table}
In order to reach larger value of $N$, the procedure then proceeds by preparing each atom successive to the third one in the superposition $\ket{\chi_k}$ of Eq.~(\ref{chik}) with relative phase
\begin{equation}\label{atomkrelativephase}
\varphi_k=\pi+\varphi_1.\quad (4\leq k\leq N)
\end{equation}
Each atom then interacts with the cavity for a time
\begin{equation}
T_k=\frac{1}{g\sqrt{k-1}}\arctan(c_{k-2}^{(k-1)}/c_{k-1}^{(k-1)}),\quad
(4\leq k\leq N),\label{Tk4}
\end{equation}
where $0\leq\arctan(c_{k-2}^{(k-1)}/c_{k-1}^{(k-1)})\leq\pi/2$. After the interaction, the cavity is in the state $\ket{\psi_{k,p,\phi}}$ of Eq.~(\ref{psikstate}) with mean phase $\phi=-\varphi_1$ and coefficients $c_n^{(k)}$, whose explicit values are given by
\begin{equation}\label{explicitc}
c_n^{(k)}=c_n^{(k-1)}\cos(g\sqrt{n}T_k)+c_{n-1}^{(k-1)}\sin(g\sqrt{n}T_k),
\end{equation}
very near to the correspondent binomial coefficients $b_n^{(k)}$. Note that the mean phase $\phi_k$ is fixed by the atomic relative phase $\varphi_1$. When $k=N$, the cavity state $\ket{\psi_{N,p,\phi}}$ of Eq.~(\ref{psikstate}) indeed represents a good approximation of the $N$GBS $\ket{N,p,\phi}$ of Eq.~(\ref{bin}). This assertion is justified by the high values of fidelity $\mathcal{F}_N(p)$. The explicit values of the coefficients $c_n^{(N)}$ for $N=3,\ldots,10$ are listed in Tab.~\ref{table}, evidencing the mismatches $\delta_n^{(N)}=[b_n^{(N)}-c_n^{(N)}]/b_n^{(N)}$ from the homologous binomial coefficients $b_n^{(N)}$ of Eq.~(\ref{bincoeff}). The comparison between $c_n^{(N)}$ and $b_n^{(N)}$ gives an accurate measure of the approximation and it can be useful in those cases where only the value of some binomial coefficients of the cavity field state is important. The mismatches are contained in the range $10^{-4}\leq\delta_n^{(N)}\leq10^{-1}$ depending on the values of $N$ and $n$. We find that the fidelity results very near to one for $4\leq N\leq10$, tends to one for $p\rightarrow0$ and in particular, for $p=1/2$, takes the value
\begin{equation}
\mathcal{F}_N(1/2)\sim1-10^{-4}.
\end{equation}
The probability $P_k(p)$ to detect each time the $k$-th atom in the ground state is found to be near to one for any value of $p$ ($P_k(p)\sim1-10^{-2}$). The probability $\mathcal{P}_N$ ($4\leq N\leq10$) to generate the cavity state $\ket{\psi_{N,p,\phi}}$ of Eq.~(\ref{psikstate}) is then calculated by Eq.~(\ref{ProbabilityN}) and it results to be inside the range
\begin{equation}
92\%\leq P_N(\ket{\downarrow_N}|\ket{\downarrow_{N-1}}|\cdots|\ket{\downarrow_1})\leq98\%,
\end{equation}
much higher than the probabilities of precedent conditional schemes ($\leq1/16$). These high generation probabilities should make our scheme of experimental interest.

In principle, we could proceed with this generation scheme for values of $N$ larger and larger, but we have to note from Eq.~(\ref{Tk4}) that the interaction time of each atom with the cavity becomes shorter and shorter. In a practical experiment, it could be difficult both to have these small interaction times and to manipulate a sequence of many atoms. A valuation of the experimental feasibility of our generation scheme is given in Sec.~\ref{experimentalfeasibility}.

\section{\label{quantumcomputation}Quantum computation with $N$GBS in CQED}
After giving an efficient method to generate $N$GBSs in a high-$Q$ cavity, in this section we study the applicative interest in the CQED framework of these states for universal quantum computation, that requires the implementation of controlled-NOT (CNOT) and 1-qubit rotation gates \cite{monroe1995PRL}. Schemes of CQED exploiting $N$GBSs to realize these universal quantum gates have been recently shown for the case $N=2$ \cite{lofranco2008IJQI}. Here we emphasize that those schemes can be immediately generalized to the case of an arbitrary value of $N$ and we treat, as an example, the realization of a CNOT gate.

\subsection{\label{dispersive}Dispersive atom-cavity interaction}
The schemes to realize the universal quantum logic gates require a Rydberg atom dispersively interacting with the cavity field state. In particular, we consider a three-level atom with energy levels $\ket{g}$, $\ket{e}$ and $\ket{i}$, where $\ket{g}$ is the ground state while $\ket{i}$ is the higher level. The cavity has a mode frequency $\omega$ slightly different from that of the transition $\ket{e}\rightarrow\ket{i}$, $\omega_{ie}$, of an amount $\delta=\omega-\omega_{ie}$ and the ground level of the atom $\ket{g}$ is unaffected by the atom-cavity coupling \cite{haroche2006book}. The coupling of the cavity mode with the atom is then with the transition $\omega_{ie}$ by the Jaynes-Cummings Hamiltonian.

If $|\delta|\gg\Omega$, where $\Omega$ is the Rabi frequency between the cavity mode and the transition $\ket{e}\rightarrow\ket{i}$, it is known that the effective atom-cavity coupling is described by the interaction Hamiltonian \cite{haroche2006book}
\begin{equation}
H_\textrm{I}=(\hbar\Omega^2/\delta)a^\dag a\sigma_{eg}^+\sigma_{eg}^-,
\end{equation}
where $a,a^\dag$ are the photon annihilation and creation operators and $\sigma_{eg}^-=\ket{g}\bra{e}$, $\sigma_{eg}^+=\ket{e}\bra{g}$. Considering an initial arbitrary cavity state $\ket{\psi}=\sum c_n\ket{n}$, the effect of this dispersive atom-cavity interaction is obtained, in interaction picture, by applying the operator $e^{-iH_\textrm{I}t/\hbar}$ to the total atom-cavity state
\begin{eqnarray}\label{effectofH}
\ket{g}\ket{\psi}&\stackrel{H_\textrm{I}}{\rightarrow}&\ket{g}\ket{\psi},\nonumber\\
\ket{e}\ket{\psi}&\stackrel{H_\textrm{I}}{\rightarrow}&\ket{e}\sum c_n e^{-in\Omega^2t/\delta}\ket{n}.
\end{eqnarray}
We now suppose the cavity initially prepared in the $N$GBS $\ket{N,1/2,\phi}$ and choose the atom-cavity interaction time $t$ such that $\Omega^2t/\delta=\pi$. Exploiting Eq.~(\ref{effectofH}) and taking into account Eq.~(\ref{bin}) together with the fact that $\ket{N,1/2,\phi-\pi}=\ket{N,1/2,\phi+\pi}$, we readily obtain
\begin{eqnarray}\label{effectofHonGBS}
\ket{g}\ket{N,1/2,\phi}&\stackrel{H_\textrm{I}}{\rightarrow}&\ket{g}\ket{N,1/2,\phi},\nonumber\\
\ket{e}\ket{N,1/2,\phi}&\stackrel{H_\textrm{I}}{\rightarrow}&\ket{e}\ket{N,1/2,\phi+\pi}.
\end{eqnarray}
We call this particular dispersive interaction (DI) the $\pi$-DI, whose effect is to leave unchanged the initial $N$GBS if the atom is in the ground state $\ket{g}$, while to transform it to its orthogonal $\ket{N,1/2,\phi+\pi}$ if the atom is in the excited state $\ket{e}$.

\subsection{Logical qubit}
We identify two orthogonal $N$GBSs as basis states of a logical qubit $\ket{0_L},\ket{1_L}$, that is
\begin{eqnarray}\label{logicalqubit}
\ket{N,1/2,\phi}&\equiv&\ket{\phi}=\ket{0_L},\nonumber\\
\ket{N,1/2,\phi+\pi}&\equiv&\ket{\phi+\pi}=\ket{1_L}.
\end{eqnarray}

An arbitrary qubit state $\ket{\psi}=a\ket{\phi}+b\ket{\phi+\pi}$, that is a quantum superposition of two orthogonal $N$GBSs, can be prepared in a cavity by the $\pi$-DI above, provided that the cavity is initially in the state $\ket{\phi}$, two opportune Ramsey zones are applied before and after the cavity and the internal atomic state is finally measured \cite{lofranco2008IJQI}. The first Ramsey zone prepares the atomic state $\ket{\chi}=a\ket{g}+b\ket{e}$ while the second Ramsey zone performs a $\pi/2$-pulse \cite{ramsey1985book}. If the outcome of the atomic measurement is $\ket{e}$, occurring with a probability of $50\%$, the procedure ends successfully. If the measurement outcome is $\ket{g}$, the qubit state obtained is instead $a\ket{\phi}-b\ket{\phi+\pi}$. We stress that this qubit preparation provides a method to generate a quantum superposition of two orthogonal $N$GBSs with $N$ arbitrarily large, generalizing therefore the generation scheme already proposed for a superposition of 2GBSs in a cavity \cite{lofranco2007PRA}.

\subsection{\label{c-not}Controlled-NOT gate with $N$GBSs}
The CNOT gate involves two qubits, namely the control qubit and the target qubit, and its action is to leave unchanged the target qubit if the control qubit is $\ket{0}$, while it flips the target qubit if the control is $\ket{1}$. The final target state can be thus written $\ket{c\oplus t}$. The global coherent action of the CNOT gate on two general qubit states $\ket{\psi_t}=a\ket{0_t}+b\ket{1_t}$ (target) and $\ket{\chi_c}=c\ket{0_c}+d\ket{1_c}$ (control) is thus
\begin{equation}\label{cnotoperation}
\ket{\chi_c}\ket{\psi_t}\stackrel{\textrm{CNOT}}{\longrightarrow}ac\ket{0_c0_t}+bc\ket{0_c1_t}+ad\ket{1_c1_t}+bd\ket{1_c0_t}.
\end{equation}

\begin{figure}
\includegraphics[height=1.90 cm, width=8.50 cm]{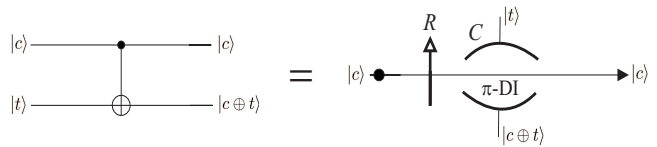}
\caption{\label{fig:CNOT}\footnotesize Scheme for realizing a controlled-NOT gate with $N$GBSs in CQED. $\ket{c}=\{\ket{g},\ket{e}\}$ is the control qubit represented by the two levels of a Rydberg atom, while $\ket{t}=\{\ket{\phi}=\ket{0_L},\ket{\phi+\pi}=\ket{1_L}\}$ is the target qubit represented by the two orthogonal $N$GBSs. $C$ indicates the cavity, $R$ the Ramsey zone and $\pi$-DI the relevant atom-cavity dispersive interaction.}
\end{figure}

In the CQED framework, this gate can be realized by the $\pi$-DI of Eq.~(\ref{effectofHonGBS}) taking the two orthogonal $N$GBSs of Eq.~(\ref{logicalqubit}) as target qubit and the atom as control qubit, with $\ket{g}=\ket{0_c}$ and $\ket{e}=\ket{1_c}$. This atom-cavity interaction, whose scheme is illustrated in Fig.~\ref{fig:CNOT}, immediately gives
\begin{eqnarray}\label{cnot2GBS}
&(c\ket{g}+d\ket{e})(a\ket{\phi}+b\ket{\phi+\pi})\stackrel{\pi\textrm{-DI}}{\longrightarrow}&\nonumber\\
&ac\ket{g,\phi}+bc\ket{g,\phi+\pi}\nonumber+ad\ket{e,\phi+\pi}+bd\ket{e,\phi},&
\end{eqnarray}
that coincides with the CNOT gate operation of Eq.~(\ref{cnotoperation}). Thus, a CNOT gate can be realized in a simple and deterministic way in CQED by exploiting $N$GBSs and an opportune dispersive interaction of a control Rydberg atom with the cavity field.

\section{\label{experimentalfeasibility}Analysis of experimental feasibility}
In this section we discuss the feasibility of the proposed generation and controlled-NOT scheme, considering the typical experimental errors involved in atom-cavity interaction systems.

Our generation scheme requires that $N$ atoms interact with the cavity each for a precise time $T_k$ ($k=1,\ldots,N$), as well as the CNOT gate scheme requires a given atom-cavity interaction time ($t=\pi\delta/\Omega^2$). However, the experimental uncertainty of the selected velocity $\Delta v$ induces an error $\Delta T$ on the interaction time such that $\Delta T/T\approx\Delta v/v$. In current laboratory experiments it is possible to obtain a relative error $\Delta v/v\leq10^{-2}$ \cite{hagley1997PRL}, which means that our schemes are rather robust to the experimental velocity fluctuations. From a comparison between the mismatches $\delta_n^{(N)}$ of Tab.~\ref{table} and the correspondent relative errors due to the experimental uncertainties $\delta_{\textrm{exp},n}^{(N)}\sim n(gT_N)^2(\Delta T_N/T_N)^2$ it is possible to show that, for the case $N=3$, the mismatches $\delta_n^{(3)}$ are smaller than $\delta_{\textrm{exp},n}^{(3)}$, while for $N\leq4$ the theoretical mismatches are of the same order or larger than the experimental ones.

We have also ignored the atomic or photon decay during the atom-cavity interactions. This assumption is valid if $\tau_\textrm{at},\tau_\textrm{cav}>T$, where $\tau_{at},\tau_{cav}$ are respectively the atom and photon mean lifetimes and $T$ is the interaction time. For Rydberg atomic levels and microwave superconducting cavities with a quality factor $10^8\leq Q\leq10^{10}$, one has $10^{-5}\textrm{s}\leq\tau_\textrm{at}\leq10^{-2}\textrm{s}$ and $10^{-4}\textrm{s}\leq\tau_\textrm{cav}\leq10^{-1}\textrm{s}$. Since typical atom-cavity field interaction times are $10^{-5}\textrm{s}\leq T\leq10^{-4}\textrm{s}$, the required condition on the mean lifetimes can be satisfied \cite{haroche2006book}. The typical lifetimes of Rydberg atomic levels, $\tau_\textrm{at}$, are such that the atoms do not decay during the sequence of the experiment and the photon mean lifetimes $\tau_\textrm{cav}$ are long enough to permit the cavity field not to decay during the interval between two consecutive atoms. The delay time between two consecutive atoms can be adjusted so that they cross the cavity one at a time, as required \cite{haroche2006book}. Moreover, the free evolution times of atoms and cavity field involved in the scheme can be determined by the geometrical configuration of the experimental apparatus.

The proposed generation scheme appears not to be affected by the non-perfect detector efficiency. Because of the high generation probability of the $N$GBS found here and being the typical atomic detection efficiency less than one ($\sim70\%\textrm{--}80\%$), the use of atomic detectors to collapse the total state to the desired cavity state cannot further reduce the uncertainty on the generated cavity state itself. So, within the experimental limits, although the generation of the desired pure $N$GBS in principle requires the final measurement of the ground atomic states, following our scheme a final atomic measurement is not required; in this sense the scheme can be considered ``quasi-deterministic''.

As said at the end of Sec.~\ref{genbin}, our procedure could be in principle extended to any value of the maximum number of photons $N$. However, the experimental capabilities limit the number of photons. For example, we have already observed that the atom-cavity interaction times become very short (tend to zero) when $N$ is very large (see Eq.~(\ref{Tk4})). Moreover, increasing the number of photons $N$ makes the decoherence time of the cavity field state to decrease as $\tau_\textrm{dec}\sim2\tau_\textrm{cav}/N$ \cite{haroche2003RoyalSoc} and this in turns limits the reachable maximum number of photons. In fact, the condition $T<\tau_\textrm{dec}$ is required in order that the cavity field does not decay during the interaction time $T$ with the atom, that is $N<2\tau_\textrm{cav}/T$. The interaction times of our scheme, given by Eq.~(\ref{Tk4}), are of the order of $T\sim1/g\sqrt{N}$ for $N\gg1$ that leads to an upper limit for $N$ given by $N<4(g\tau_\textrm{cav})^2$. For a typical Fabry-Perot cavity technology ($\tau_\textrm{cav}=10^{-3}\ \textrm{s}$, $g=2\pi\times50\ \textrm{kHz}$ \cite{haroche2003RoyalSoc}) this inequality gives $N<10^4$. Currently, it is possible to manipulate sequences of few atoms ($N<10$). However, the CQED experimental developments are rather promising and it could be soon possible to control a sequence of more than three atoms and generate a $N$GBS in laboratory \cite{haroche2006book,maioli2005PRL}.

\section{\label{conclu}Conclusive remarks}
In this paper we have proposed an efficient scheme to generate $N$-photon generalized binomial states ($N$GBSs) with a maximum number of photons $N>2$ in the CQED framework. The scheme utilizes $N$ consecutive two-level atoms interacting, one at a time, with the cavity field initially in the vacuum state. We have shown that, suitably fixing the interaction time of each atom with the cavity, the atom exits the cavity in the ground state with a probability near to one and, after the passage of the $N$-th atom, the resulting cavity field state is, with high fidelity, a $N$GBS. The parameters of the generated $N$GBS, as probability of single photon occurrence $p$ and mean phase $\phi$, are directly connected to the settings of the Ramsey zone that prepares the atomic state superposition before the atom enters the cavity (Sec.~\ref{genbin}). We have also discuss the applicative interest of $N$GBSs in quantum computation processing, showing that universal quantum gates can be realized in the CQED framework by choosing two orthogonal $N$GBSs as the logical qubit stored inside the cavity and exploiting an opportune dispersive atom-cavity interaction. In particular, we have described the realization of a CNOT gate using a Rydberg atom as control qubit (Sec.~\ref{quantumcomputation}). From an applicative point of view, besides quantum computation, one could utilize the generation scheme proposed here to produce the $N$GBS $\ket{N,1/2,\pi}$, having $p=1/2$ and $\phi=\pi$, that is exploited as reference field state for measuring the canonical phase of quantum electromagnetic fields \cite{pregnell2002PRL,pregnell2003PRA}. We have finally analyzed the feasibility of the proposed schemes, that seem not to be sensibly affected by the typical experimental limitations, in particular for values of $N\sim10$. We have pointed out that the maximum number of photons $N$ reachable for the $N$GBSs generated by our efficient scheme is limited only by the current experimental capabilities (Sec.~\ref{experimentalfeasibility}).

The results of this paper open the way to the realization of schemes aimed at generating quantum superpositions of two orthogonal $N$GBSs inside a cavity (see, for example, the end of Sec.~\ref{dispersive}) or entanglement of $N$GBSs in two spatially separate cavities, which can be useful both for investigations of the foundations of quantum theory, as classical-quantum border or nonlocal properties, and for quantum information processing. Moreover, the implementation of quantum computation protocols exploiting $N$GBSs and Rydberg atoms in CQED as suggested here (see Sec.~\ref{quantumcomputation}) appears to be more convenient than that based on coherent states, where the non exact orthogonality and the requirement of teleportation protocols constitute some drawbacks \cite{jeong2002PRA}.

\begin{acknowledgments}
G.C. (A.M.) acknowledges partial support by MIUR project II04C0E3F3 (II04C1AF4E) \textit{Collaborazioni Interuniversitarie ed Internazionali tipologia C}.
\end{acknowledgments}

\appendix
\section{\label{binomialcoeffcondition}Proof of the binomial coefficients condition}
Let us suppose that, after fixing $\Phi_k=0$ in Eq.~(\ref{coefficientsaandc}), the conditions
\begin{equation}
b_{n-1}^{(k-1)}\cos(g\sqrt{n}T_k)+b_{n}^{(k-1)}\sin(g\sqrt{n}T_k)=0,\label{eqcoeffbis}
\end{equation}
are simultaneously satisfied for $n=1,\ldots,k$. If one takes negative values of the functions $\sin(g\sqrt{n}T_k)$, the previous conditions give
\begin{eqnarray}
\sin(g\sqrt{n}T_k)&=&-\frac{b_{n-1}^{(k-1)}}{\sqrt{(b_{n-1}^{(k-1)})^2+(b_{n}^{(k-1)})^2}},\nonumber\\
\cos(g\sqrt{n}T_k)&=&\frac{b_{n}^{(k-1)}}{\sqrt{(b_{n-1}^{(k-1)})^2+(b_{n}^{(k-1)})^2}}.\label{sincos}
\end{eqnarray}
Using the definition of binomial coefficients given in Eq.~(\ref{bincoeff}), from Eq.~(\ref{sincos}) we then obtain
\begin{eqnarray}
&&b_{n}^{(k-1)}\cos(g\sqrt{n}T_k)-b_{n-1}^{(k-1)}\sin(g\sqrt{n}T_k)\nonumber\\
&&=\sqrt{(b_{n-1}^{(k-1)})^2+(b_{n}^{(k-1)})^2}\nonumber\\
&&=\sqrt{\frac{(k-1)!}{(k-n)!(n-1)!}+\frac{(k-1)!}{(k-n-1)!n!}}\nonumber\\
&&=\sqrt{\frac{k!}{(k-n)!n!}\left(\frac{n+k-n}{k}\right)},
\end{eqnarray}
and finally
\begin{equation}\label{binomialconditions}
b_{n}^{(k-1)}\cos(g\sqrt{n}T_k)-b_{n-1}^{(k-1)}\sin(g\sqrt{n}T_k)=b_n^{(k)}.
\end{equation}
These are just the required conditions, $c_n^{(k)}=b_n^{(k)}$, in order that the total atom-cavity state resulting from Eq.~(\ref{Psik}) is factorized as $\ket{k,p,\phi}\ket{\downarrow}$, where $\ket{k,p,\phi}$ is the $N$GBS of Eq.~(\ref{bin}) with $N=k$. Note that if one fixes $\Phi_k=\pi$ in Eq.~(\ref{coefficientsaandc}) the sign ``+'' of Eq.~(\ref{eqcoeffbis}) must be replaced by ``$-$'' and we would obtain a result analogous to Eq.~(\ref{binomialconditions}), with the sign ``+'' in place of ``$-$'', by considering positive values of the trigonometric functions of Eq.~(\ref{sincos}).


\begin{thebibliography}{32}
\expandafter\ifx\csname natexlab\endcsname\relax\def\natexlab#1{#1}\fi
\expandafter\ifx\csname bibnamefont\endcsname\relax
  \def\bibnamefont#1{#1}\fi
\expandafter\ifx\csname bibfnamefont\endcsname\relax
  \def\bibfnamefont#1{#1}\fi
\expandafter\ifx\csname citenamefont\endcsname\relax
  \def\citenamefont#1{#1}\fi
\expandafter\ifx\csname url\endcsname\relax
  \def\url#1{\texttt{#1}}\fi
\expandafter\ifx\csname urlprefix\endcsname\relax\def\urlprefix{URL }\fi
\providecommand{\bibinfo}[2]{#2}
\providecommand{\eprint}[2][]{\url{#2}}

\bibitem[{\citenamefont{Haroche and Raimond}(2006)}]{haroche2006book}
\bibinfo{author}{\bibfnamefont{S.}~\bibnamefont{Haroche}} \bibnamefont{and}
  \bibinfo{author}{\bibfnamefont{J.~M.} \bibnamefont{Raimond}},
  \emph{\bibinfo{title}{Exploring the Quantum: Atoms, Cavities, and Photons}}
  (\bibinfo{publisher}{Oxford University Press, USA}, \bibinfo{address}{Oxford,
  New York}, \bibinfo{year}{2006}).

\bibitem[{\citenamefont{Nielsen and Chuang}(2000)}]{nielsenchuang}
\bibinfo{author}{\bibfnamefont{M.~A.} \bibnamefont{Nielsen}} \bibnamefont{and}
  \bibinfo{author}{\bibfnamefont{I.~L.} \bibnamefont{Chuang}},
  \emph{\bibinfo{title}{Quantum Computation and Quantum Information}}
  (\bibinfo{publisher}{Cambridge University Press}, \bibinfo{year}{2000}).

\bibitem[{\citenamefont{Zeilinger}(1998)}]{zeilinger1998RevModPhys}
\bibinfo{author}{\bibfnamefont{A.}~\bibnamefont{Zeilinger}},
  \bibinfo{journal}{Rev. Mod. Phys.} \textbf{\bibinfo{volume}{71}},
  \bibinfo{pages}{S288} (\bibinfo{year}{1998}).

\bibitem[{\citenamefont{Raimond et~al.}(2001)\citenamefont{Raimond, Brune, and
  Haroche}}]{raimond2001RevModPhys}
\bibinfo{author}{\bibfnamefont{J.~M.} \bibnamefont{Raimond}},
  \bibinfo{author}{\bibfnamefont{M.}~\bibnamefont{Brune}}, \bibnamefont{and}
  \bibinfo{author}{\bibfnamefont{S.}~\bibnamefont{Haroche}},
  \bibinfo{journal}{Rev. Mod. Phys.} \textbf{\bibinfo{volume}{73}},
  \bibinfo{pages}{565} (\bibinfo{year}{2001}).

\bibitem[{\citenamefont{Meystre}(1992)}]{meystre}
\bibinfo{author}{\bibfnamefont{P.}~\bibnamefont{Meystre}}, in
  \emph{\bibinfo{booktitle}{Progress in Optics XXX, Cavity Quantum Optics and
  the Quantum Measurement Process}} (\bibinfo{publisher}{Elsevier Science
  Publishers B.V.}, \bibinfo{address}{New York}, \bibinfo{year}{1992}).

\bibitem[{\citenamefont{Vogel et~al.}(1993)\citenamefont{Vogel, Akulin, and
  Schleich}}]{vogel1993PRL}
\bibinfo{author}{\bibfnamefont{K.}~\bibnamefont{Vogel}},
  \bibinfo{author}{\bibfnamefont{V.~M.} \bibnamefont{Akulin}},
  \bibnamefont{and} \bibinfo{author}{\bibfnamefont{W.~P.}
  \bibnamefont{Schleich}}, \bibinfo{journal}{Phys. Rev. Lett.}
  \textbf{\bibinfo{volume}{71}}, \bibinfo{pages}{1816} (\bibinfo{year}{1993}).

\bibitem[{\citenamefont{Bertet et~al.}(2002)}]{bertet2002PRL}
\bibinfo{author}{\bibfnamefont{P.}~\bibnamefont{Bertet}} \bibnamefont{et~al.},
  \bibinfo{journal}{Phys. Rev. Lett.} \textbf{\bibinfo{volume}{88}},
  \bibinfo{pages}{143601} (\bibinfo{year}{2002}).

\bibitem[{\citenamefont{Stoler et~al.}(1985)\citenamefont{Stoler, Saleh, and
  Teich}}]{stoler1985OptActa}
\bibinfo{author}{\bibfnamefont{D.}~\bibnamefont{Stoler}},
  \bibinfo{author}{\bibfnamefont{B.~E.~A.} \bibnamefont{Saleh}},
  \bibnamefont{and} \bibinfo{author}{\bibfnamefont{M.~C.} \bibnamefont{Teich}},
  \bibinfo{journal}{Opt. Acta} \textbf{\bibinfo{volume}{32}},
  \bibinfo{pages}{345} (\bibinfo{year}{1985}).

\bibitem[{\citenamefont{Vidiella-Barranco and Roversi}(1994)}]{vidiella1994PRA}
\bibinfo{author}{\bibfnamefont{A.}~\bibnamefont{Vidiella-Barranco}}
  \bibnamefont{and} \bibinfo{author}{\bibfnamefont{J.~A.}
  \bibnamefont{Roversi}}, \bibinfo{journal}{Phys. Rev. A}
  \textbf{\bibinfo{volume}{50}}, \bibinfo{pages}{5233} (\bibinfo{year}{1994}).

\bibitem[{\citenamefont{Pregnell and Pegg}(2002)}]{pregnell2002PRL}
\bibinfo{author}{\bibfnamefont{K.~L.} \bibnamefont{Pregnell}} \bibnamefont{and}
  \bibinfo{author}{\bibfnamefont{D.~T.} \bibnamefont{Pegg}},
  \bibinfo{journal}{Phys. Rev. Lett.} \textbf{\bibinfo{volume}{89}},
  \bibinfo{pages}{173601} (\bibinfo{year}{2002}).

\bibitem[{\citenamefont{Pregnell and Pegg}(2003)}]{pregnell2003PRA}
\bibinfo{author}{\bibfnamefont{K.~L.} \bibnamefont{Pregnell}} \bibnamefont{and}
  \bibinfo{author}{\bibfnamefont{D.~T.} \bibnamefont{Pegg}},
  \bibinfo{journal}{Phys. Rev. A} \textbf{\bibinfo{volume}{67}},
  \bibinfo{pages}{063814} (\bibinfo{year}{2003}).

\bibitem[{\citenamefont{{Lo Franco} et~al.}(2008)\citenamefont{{Lo Franco},
  Compagno, Messina, and Napoli}}]{lofranco2008EPJ}
\bibinfo{author}{\bibfnamefont{R.}~\bibnamefont{{Lo Franco}}},
  \bibinfo{author}{\bibfnamefont{G.}~\bibnamefont{Compagno}},
  \bibinfo{author}{\bibfnamefont{A.}~\bibnamefont{Messina}}, \bibnamefont{and}
  \bibinfo{author}{\bibfnamefont{A.}~\bibnamefont{Napoli}},
  \bibinfo{journal}{Eur. Phys. J. ST} \textbf{\bibinfo{volume}{160}},
  \bibinfo{pages}{247} (\bibinfo{year}{2008}).

\bibitem[{\citenamefont{Arecchi et~al.}(1972)\citenamefont{Arecchi, Courtens,
  Gilmoure, and Thomas}}]{arecchi1972PRA}
\bibinfo{author}{\bibfnamefont{F.~T.} \bibnamefont{Arecchi}},
  \bibinfo{author}{\bibfnamefont{E.}~\bibnamefont{Courtens}},
  \bibinfo{author}{\bibfnamefont{R.}~\bibnamefont{Gilmoure}}, \bibnamefont{and}
  \bibinfo{author}{\bibfnamefont{H.}~\bibnamefont{Thomas}},
  \bibinfo{journal}{Phys. Rev. A} \textbf{\bibinfo{volume}{6}},
  \bibinfo{pages}{2211} (\bibinfo{year}{1972}).

\bibitem[{\citenamefont{{Lo Franco}
  et~al.}(2006{\natexlab{a}})\citenamefont{{Lo Franco}, Compagno, Messina, and
  Napoli}}]{lofranco2006PRA}
\bibinfo{author}{\bibfnamefont{R.}~\bibnamefont{{Lo Franco}}},
  \bibinfo{author}{\bibfnamefont{G.}~\bibnamefont{Compagno}},
  \bibinfo{author}{\bibfnamefont{A.}~\bibnamefont{Messina}}, \bibnamefont{and}
  \bibinfo{author}{\bibfnamefont{A.}~\bibnamefont{Napoli}},
  \bibinfo{journal}{Phys. Rev. A} \textbf{\bibinfo{volume}{74}},
  \bibinfo{pages}{045803} (\bibinfo{year}{2006}{\natexlab{a}}).

\bibitem[{\citenamefont{{Lo Franco}
  et~al.}(2007{\natexlab{a}})\citenamefont{{Lo Franco}, Compagno, Messina, and
  Napoli}}]{lofranco2007PRA}
\bibinfo{author}{\bibfnamefont{R.}~\bibnamefont{{Lo Franco}}},
  \bibinfo{author}{\bibfnamefont{G.}~\bibnamefont{Compagno}},
  \bibinfo{author}{\bibfnamefont{A.}~\bibnamefont{Messina}}, \bibnamefont{and}
  \bibinfo{author}{\bibfnamefont{A.}~\bibnamefont{Napoli}},
  \bibinfo{journal}{Phys. Rev. A} \textbf{\bibinfo{volume}{76}},
  \bibinfo{pages}{011804(R)} (\bibinfo{year}{2007}{\natexlab{a}}).

\bibitem[{\citenamefont{{Lo Franco}
  et~al.}(2006{\natexlab{b}})\citenamefont{{Lo Franco}, Compagno, Messina, and
  Napoli}}]{lofranco2006OpenSys}
\bibinfo{author}{\bibfnamefont{R.}~\bibnamefont{{Lo Franco}}},
  \bibinfo{author}{\bibfnamefont{G.}~\bibnamefont{Compagno}},
  \bibinfo{author}{\bibfnamefont{A.}~\bibnamefont{Messina}}, \bibnamefont{and}
  \bibinfo{author}{\bibfnamefont{A.}~\bibnamefont{Napoli}},
  \bibinfo{journal}{Open Sys. and Information Dyn.}
  \textbf{\bibinfo{volume}{13}}, \bibinfo{pages}{463}
  (\bibinfo{year}{2006}{\natexlab{b}}).

\bibitem[{\citenamefont{{Lo Franco}
  et~al.}(2007{\natexlab{b}})\citenamefont{{Lo Franco}, Compagno, Messina, and
  Napoli}}]{lofranco2007OptSpect}
\bibinfo{author}{\bibfnamefont{R.}~\bibnamefont{{Lo Franco}}},
  \bibinfo{author}{\bibfnamefont{G.}~\bibnamefont{Compagno}},
  \bibinfo{author}{\bibfnamefont{A.}~\bibnamefont{Messina}}, \bibnamefont{and}
  \bibinfo{author}{\bibfnamefont{A.}~\bibnamefont{Napoli}},
  \bibinfo{journal}{Opt. Spectrosc.} \textbf{\bibinfo{volume}{103}},
  \bibinfo{pages}{890} (\bibinfo{year}{2007}{\natexlab{b}}).

\bibitem[{\citenamefont{{Lo Franco} et~al.}(2005)\citenamefont{{Lo Franco},
  Compagno, Messina, and Napoli}}]{lofranco2005PRA}
\bibinfo{author}{\bibfnamefont{R.}~\bibnamefont{{Lo Franco}}},
  \bibinfo{author}{\bibfnamefont{G.}~\bibnamefont{Compagno}},
  \bibinfo{author}{\bibfnamefont{A.}~\bibnamefont{Messina}}, \bibnamefont{and}
  \bibinfo{author}{\bibfnamefont{A.}~\bibnamefont{Napoli}},
  \bibinfo{journal}{Phys. Rev. A} \textbf{\bibinfo{volume}{72}},
  \bibinfo{pages}{053806} (\bibinfo{year}{2005}).

\bibitem[{\citenamefont{Dattoli et~al.}(1987)\citenamefont{Dattoli, Gallardo,
  and Torre}}]{dattoli1987JOptSoc}
\bibinfo{author}{\bibfnamefont{G.}~\bibnamefont{Dattoli}},
  \bibinfo{author}{\bibfnamefont{J.}~\bibnamefont{Gallardo}}, \bibnamefont{and}
  \bibinfo{author}{\bibfnamefont{A.}~\bibnamefont{Torre}}, \bibinfo{journal}{J.
  Opt. Soc. of Am. B} \textbf{\bibinfo{volume}{2}}, \bibinfo{pages}{185}
  (\bibinfo{year}{1987}).

\bibitem[{\citenamefont{Vidiella-Barranco and
  Roversi}(1995)}]{vidiella1995JModOpt}
\bibinfo{author}{\bibfnamefont{A.}~\bibnamefont{Vidiella-Barranco}}
  \bibnamefont{and} \bibinfo{author}{\bibfnamefont{J.~A.}
  \bibnamefont{Roversi}}, \bibinfo{journal}{J. Mod. Optics}
  \textbf{\bibinfo{volume}{42}}, \bibinfo{pages}{2475} (\bibinfo{year}{1995}).

\bibitem[{\citenamefont{Joshi and Puri}(1989{\natexlab{a}})}]{jos1}
\bibinfo{author}{\bibfnamefont{A.}~\bibnamefont{Joshi}} \bibnamefont{and}
  \bibinfo{author}{\bibfnamefont{R.~R.} \bibnamefont{Puri}},
  \bibinfo{journal}{J. Mod. Optics} \textbf{\bibinfo{volume}{36}},
  \bibinfo{pages}{215} (\bibinfo{year}{1989}{\natexlab{a}}).

\bibitem[{\citenamefont{Joshi and Puri}(1989{\natexlab{b}})}]{jos2}
\bibinfo{author}{\bibfnamefont{A.}~\bibnamefont{Joshi}} \bibnamefont{and}
  \bibinfo{author}{\bibfnamefont{R.~R.} \bibnamefont{Puri}},
  \bibinfo{journal}{J. Mod. Optics} \textbf{\bibinfo{volume}{36}},
  \bibinfo{pages}{557} (\bibinfo{year}{1989}{\natexlab{b}}).

\bibitem[{\citenamefont{Jaynes and Cummings}(1963)}]{jaynescummings}
\bibinfo{author}{\bibfnamefont{E.~T.} \bibnamefont{Jaynes}} \bibnamefont{and}
  \bibinfo{author}{\bibfnamefont{F.~W.} \bibnamefont{Cummings}},
  \bibinfo{journal}{P.I.E.E.E.} \textbf{\bibinfo{volume}{51}},
  \bibinfo{pages}{89} (\bibinfo{year}{1963}).

\bibitem[{\citenamefont{Scully and Zubairy}(1997)}]{scully1997book}
\bibinfo{author}{\bibfnamefont{M.~O.} \bibnamefont{Scully}} \bibnamefont{and}
  \bibinfo{author}{\bibfnamefont{M.~S.} \bibnamefont{Zubairy}},
  \emph{\bibinfo{title}{Quantum Optics}} (\bibinfo{publisher}{Cambridge
  University Press}, \bibinfo{year}{1997}).

\bibitem[{\citenamefont{Moussa and Baseia}(1998)}]{moussa1998PLA}
\bibinfo{author}{\bibfnamefont{M.}~\bibnamefont{Moussa}} \bibnamefont{and}
  \bibinfo{author}{\bibfnamefont{B.}~\bibnamefont{Baseia}},
  \bibinfo{journal}{Phys. Lett. A} \textbf{\bibinfo{volume}{238}},
  \bibinfo{pages}{223} (\bibinfo{year}{1998}).

\bibitem[{\citenamefont{Ramsey}(1985)}]{ramsey1985book}
\bibinfo{author}{\bibfnamefont{N.~F.} \bibnamefont{Ramsey}},
  \emph{\bibinfo{title}{Molecular Beams}} (\bibinfo{publisher}{Oxford
  University Press}, \bibinfo{year}{1985}).

\bibitem[{\citenamefont{Hagley et~al.}(1997)}]{hagley1997PRL}
\bibinfo{author}{\bibfnamefont{E.}~\bibnamefont{Hagley}} \bibnamefont{et~al.},
  \bibinfo{journal}{Phys. Rev. Lett.} \textbf{\bibinfo{volume}{79}},
  \bibinfo{pages}{1} (\bibinfo{year}{1997}).

\bibitem[{\citenamefont{Monroe et~al.}(1995)\citenamefont{Monroe, Meekhof,
  King, Itano, and Wineland}}]{monroe1995PRL}
\bibinfo{author}{\bibfnamefont{C.}~\bibnamefont{Monroe}},
  \bibinfo{author}{\bibfnamefont{D.~M.} \bibnamefont{Meekhof}},
  \bibinfo{author}{\bibfnamefont{B.~E.} \bibnamefont{King}},
  \bibinfo{author}{\bibfnamefont{W.~M.} \bibnamefont{Itano}}, \bibnamefont{and}
  \bibinfo{author}{\bibfnamefont{D.~J.} \bibnamefont{Wineland}},
  \bibinfo{journal}{Phys. Rev. Lett.} \textbf{\bibinfo{volume}{75}},
  \bibinfo{pages}{4714} (\bibinfo{year}{1995}).

\bibitem[{\citenamefont{{Lo Franco} et~al.}()\citenamefont{{Lo Franco},
  Compagno, Messina, and Napoli}}]{lofranco2008IJQI}
\bibinfo{author}{\bibfnamefont{R.}~\bibnamefont{{Lo Franco}}},
  \bibinfo{author}{\bibfnamefont{G.}~\bibnamefont{Compagno}},
  \bibinfo{author}{\bibfnamefont{A.}~\bibnamefont{Messina}}, \bibnamefont{and}
  \bibinfo{author}{\bibfnamefont{A.}~\bibnamefont{Napoli}},
  \bibinfo{note}{preprint quant-ph/0805.2282}.

\bibitem[{\citenamefont{Haroche}(2003)}]{haroche2003RoyalSoc}
\bibinfo{author}{\bibfnamefont{S.}~\bibnamefont{Haroche}},
  \bibinfo{journal}{Phil. Trans. R. Soc. Lond. A}
  \textbf{\bibinfo{volume}{361}}, \bibinfo{pages}{1339} (\bibinfo{year}{2003}).

\bibitem[{\citenamefont{Maioli et~al.}(2005)}]{maioli2005PRL}
\bibinfo{author}{\bibfnamefont{P.}~\bibnamefont{Maioli}} \bibnamefont{et~al.},
  \bibinfo{journal}{Phys. Rev. Lett.} \textbf{\bibinfo{volume}{94}},
  \bibinfo{pages}{113601} (\bibinfo{year}{2005}).

\bibitem[{\citenamefont{Jeong and Kim}(2002)}]{jeong2002PRA}
\bibinfo{author}{\bibfnamefont{H.}~\bibnamefont{Jeong}} \bibnamefont{and}
  \bibinfo{author}{\bibfnamefont{M.~S.} \bibnamefont{Kim}},
  \bibinfo{journal}{Phys. Rev. A} \textbf{\bibinfo{volume}{65}},
  \bibinfo{pages}{042305} (\bibinfo{year}{2002}).

\end{thebibliography}
\end{document}